\documentclass[twocolumn,showpacs,preprintnumbers,amsmath,amssymb,pra,floatfix]{revtex4}
\usepackage{graphicx}% Include figure files
\usepackage{dcolumn}% Align table columns on decimal point
\usepackage{bm}% bold math
\usepackage{epsfig}
\usepackage{bm}
\newcommand{\beeq}{\begin{equation}}
\newcommand{\eneq}{\end{equation}}
\newcommand{\bear}{\begin{array}}
\newcommand{\enar}{\end{array}}
\newcommand{\fone}{\mathbf{S}_1}
\newcommand{\ftwo}{\mathbf{S}_2}
\newcommand{\F}{\mathbf{S}}
\newcommand{\rr}{\mathbf{r}}
\newcommand{\f}{\mathbf{S}}

\begin{document}

\title{Ground states of spin-3 Bose-Einstein condensates for conserved 
magnetization}

\author{H. M\"akel\"a and K.-A. Suominen}
% \altaffiliation[Also at ]{Physics Department, XYZ University.}
 %x\email{Second.Author@institution.edu}
\affiliation{Department of Physics, University of Turku, 
FIN-20014 Turun yliopisto, Finland
}%

\date{\today}

\begin{abstract}
We calculate the ground states and ground state phase diagrams of Bose-Einstein 
condensates of spin-3 atoms under the assumption of conserved magnetization.   
We especially concentrate on the ground states of a ${}^{52}$Cr condensate. 
In ${}^{52}$Cr the magnetic dipole-dipole interaction enables magnetization 
changing collisions, but in a strong magnetic field these are suppressed. 
In the calculation of the phase diagrams we neglect the contribution from the 
dipole-dipole interaction, but discuss its effects at the end of the paper. 
We show that the ground state of a ${}^{52}$Cr condensate does not seem to depend on  
whether or not the dipole-dipole interaction is taken into attention.  
\end{abstract}

\maketitle                              %display desired

\section{INTRODUCTION} 

Interest in spin-3 Bose-Einstein condensates (BECs) has  
increased rapidly since the creation of a chromium condensate \cite{Griesmaier05}.  
In a chromium condensate the magnetic dipole-dipole interaction is much stronger 
than in alkali-metal atom condensates. Therefore   
two interactions need to be considered:  
the short-range (van der Waals) interaction modeled as a contact interaction, 
and the long-range magnetic  
dipole-dipole interaction.  
The latter is known to enable conversion 
of spin angular momentum into orbital angular momentum, which    
makes possible the creation of vortex states via spin dynamics \cite{Kawaguchi05,Santos05}. 
The spin dynamics driven by the 
dipole-dipole interaction can be prevented by exposing the condensate to a strong external magnetic field, 
which suppresses the conversion of spin angular momentum into orbital angular momentum  
\cite{Santos05,Kawaguchi05,Kawaguchi06}.   
The dipole-dipole interaction also affects the expansion dynamics of 
a chromium condensate \cite{Stuhler05,Giovanazzi06a}, allows a precise   
 measurement of the scattering length \cite{Griesmaier06,Giovanazzi06b}, and 
opens the way for a new cooling method \cite{Fattori06}.  
One difference between ${}^{52}$Cr and alkali-metal atoms is that in the former 
the nuclear spin is zero. Therefore the quadratic 
Zeeman effect is absent when a chromium condensate is placed in   
 a magnetic field. Recently it was shown that by using laser fields an effective 
quadratic Zeeman effect can be created also in  ${}^{52}$Cr \cite{Santos06}. 
  
In experiments performed so far, the condensate has been prepared in 
the $S=3$, $m_S=-3$ state, where $S$ is the total electronic spin 
and $m_S$ gives the projection of $S$ in the direction of the magnetic field. 
In this state the effect of the dipole-dipole interaction is the largest, but even then it is  
weaker than the van der Waals interaction. Therefore it has been neglected 
in the studies of the ground states of a ${}^{52}$Cr condensate \cite{Santos05,Diener05}. 
This simplifies the study of the system, 
as it allows one to concentrate on the effects of the contact interaction alone.  
Since the contact interaction does not change the value of magnetization, the 
ground states should be calculated taking this into attention. 
For spin-1 condensate this has been done in Ref. \cite{Zhang03}.

In this paper we calculate the ground states of spin-3 condensates 
 under the assumption of conserved magnetization. The ground states 
of spin-3 condensates have been 
studied earlier, but in the previous studies the magnetization has been allowed to 
change freely \cite{Santos05,Diener05,Barnett06,Barnett06b,Yip06}.    
We also present the ground-state phase diagrams of spin-3 condensates corresponding to several 
fixed values of the magnetization. Finally we  
discuss the special case of ${}^{52}$Cr and study qualitatively the role 
of the magnetic dipole-dipole interaction.

This paper is organized as follows. In Sec. \ref{Sec2} we derive an expression 
for the ground-state energy of a spin-3 condensate. 
In Sec. \ref{Sec3} we calculate the ground states and the ground state  
phase diagrams. In  
Sec. \ref{Sec4} we study the effects of the dipole-dipole interaction, analyze  
the ground states of ${}^{52}$Cr, and discuss the possible experiments. 
Section \ref{Sec5} contains a summary of the results of the paper.

\section{\label{Sec2}THE INTERACTION ENERGY}
In this section we derive an expression for the energy of a spin-3 condensate. 
The contact interaction between two condensate atoms can be 
 written as  
$U(\mathbf{r}-\mathbf{r}')=\delta(\mathbf{r}-\mathbf{r}')
\sum_{s=0}^{6} g_{s}\mathcal{P}_{s}$, where $\mathcal{P}_{s}$ 
is the projection operator to a two-particle state with total spin
$s$, $g_{s}=\frac{4\pi\hbar^2 a_{s}}{M}$ and $a_{s}$ is the $s$-wave
scattering length in the total spin $s$ channel. 
Because of symmetry, only even terms appear in the sum \cite{Ho98}.   
We consider a two-particle system and  
denote by $\f_i$ the spin operator of particle $i$, that is, 
$\f_1=\F\otimes I_2$,  $\f_2=I_1\otimes\F$  and $\F$ is the spin
 operator of a spin-$3$ particle.  
The total spin operator is $\F_{tot}=\fone+\ftwo$. 
The possible values for the total spin of two particles are now  
$0,2,4,6$ and therefore we get     
$\F_{tot}^2=\sum_{s=0,2,4,6} s(s+1)\mathcal{P}_{s}=6\mathcal{P}_2
+20\mathcal{P}_4+42\mathcal{P}_6$ $(\hbar=1)$.   
The sum of the projection operators gives the identity operator,  
$I\equiv I_1\otimes I_2=
\mathcal{P}_{0}+\mathcal{P}_{2}+\mathcal{P}_{4}+\mathcal{P}_{6}$.
Now 
$\fone\cdot \ftwo=(\F_{tot}^2-\F_1^2-\F_2^2)/2=
(\F_{tot}^2-24 I)/2=-12\mathcal{P}_0 -9\mathcal{P}_2
-2\mathcal{P}_4+9\mathcal{P}_6$. By squaring this 
we get $(\fone\cdot \ftwo)^2=144\mathcal{P}_0 
+81\mathcal{P}_2+4\mathcal{P}_4+81\mathcal{P}_6=81I+63\mathcal{P}_0 -77\mathcal{P}_4 $. 
The latter expression is obtained by using the equation 
$\mathcal{P}_6 =I-\mathcal{P}_0 -\mathcal{P}_2-\mathcal{P}_4$.   

The projection
 operator $\mathcal{P}_0$ can be written as
 $\mathcal{P}_0=|00\rangle\langle 00|$, where 
 $|00\rangle\equiv |S_{tot}=0,m_{S_{tot}}=0\rangle$.  
With the help of the expressions for $I,\mathcal{P}_0, 
\fone\cdot \ftwo$, and $(\fone\cdot \ftwo)^2$ we can write 
 $\mathcal{P}_2,\mathcal{P}_4$ and $\mathcal{P}_6$ as  
$\mathcal{P}_2=-\frac{1}{126}[18 I+210\mathcal{P}_0
+7\fone\cdot \ftwo-(\fone\cdot \ftwo)^2],
\mathcal{P}_4=\frac{1}{77}[81 I+63\mathcal{P}_0
-(\fone\cdot  \ftwo)^2]$ and $\mathcal{P}_6=\frac{1}{198}[18 I
-30\mathcal{P}_0 +11\fone\cdot \ftwo  
+(\fone\cdot \ftwo)^2]$. 
Using these the interaction term can be written as 
\begin{equation}
U(\mathbf{r}-\mathbf{r}')=
\delta(\mathbf{r}-\mathbf{r}')[\alpha I+7\beta\mathcal{P}_0
+\gamma\fone\cdot\ftwo  
+\delta(\fone\cdot \ftwo)^2],
\end{equation} 
where 
$\alpha=-\frac{1}{7}g_2+\frac{81}{77}g_4+\frac{1}{11}g_6$, 
$7\beta=g_0-\frac{5}{3}g_2+\frac{9}{11}g_4-\frac{5}{33}g_6$, 
$\gamma=\frac{1}{18}(g_6-g_2)$ and 
$\delta=\frac{1}{126}g_2-\frac{1}{77}g_4+\frac{1}{198}g_6$. 
The measured values for the scattering lengths of ${}^{52}$Cr are   
$a_2=-7a_B,a_4=58a_B,
a_6=112a_B$, while the value of $a_0$ is unknown ($a_B$ is the Bohr
radius) \cite{Werner05}. We write  
$\alpha=\frac{4\pi\hbar^2 a_B}{M}\alpha'$,   
and define $\beta',\gamma',\delta'$ similarly. The scattering lengths give 
$\alpha'=72.19a_B,7\beta'=42.15 a_B+a_0,\gamma'=6.61a_B,\delta'=-0.24 a_B$.

Next we study the energy of a spin-3 condensate using mean-field theory. 
We write $\psi=\sqrt{n}\xi$, where $n$ is the particle density and 
$\xi$ is a complex vector with seven 
components and $\xi^\dag \xi=1$. We call $\xi$ a spinor.  
Previous studies 
have shown that the contribution from the magnetic dipole-dipole interaction
is much smaller than that of the contact interaction \cite{Santos05,Diener05}. 
Here we also neglect the magnetic dipole-dipole interaction term at first, 
but discuss its effects in Sec. \ref{Sec4}. If the dipole-dipole interaction 
is neglected the Hamiltonian reads  
\begin{eqnarray}
H=H_0+
\int d\rr\,\frac{n^2}{2}\bigg\{\alpha+\xi_\mu^*\xi_\tau^*\bigg[7\beta\,\langle
3\mu;3\tau|00\rangle\langle 00|3\mu';3\tau'\rangle \nonumber\\*
 +\,\gamma\,
\f_{1\mu\mu'}\cdot\f_{2\tau\tau'}+\delta\sum_{ij}(S_i S_j)_{\mu\mu'}
(S_i S_j)_{\tau\tau'}\bigg]\xi_{\tau'}\xi_{\mu'}\bigg\}. 
\end{eqnarray} 
Here $H_0=\int d\rr[\hbar^2/(2M)\nabla\psi^*_\mu\cdot
\nabla\psi_\mu +
Un]$ with $U$ an external potential, repeated index is
summed, $i,j=x,y,z$,  and $\f_{1\mu\mu'}=\f_{\mu\mu'}
\otimes I_2$, 
$\f_{2\tau\tau'}=I_1\otimes \f_{\tau\tau'}$. 

The state $|00\rangle$ appearing in the 
projection operator $\mathcal{P}_0$ 
can be written in terms of one-particle states as $|00\rangle=
\frac{1}{\sqrt{7}}[|3-3;33\rangle +|33;3-3\rangle
-|3-2;32\rangle -|32;3-2\rangle +|3-1;31\rangle 
+|31;3-1\rangle -|30;30\rangle]$. 
Using this the energy becomes 
%\begin{widetext}
\begin{eqnarray}\label{E}
E[\psi]&=&\int d\rr\bigg(\frac{\hbar^2}{2m}\nabla\psi^*_\mu
\nabla\psi_\mu + Un+\frac{n^2}{2}\nonumber\\*
&&\times\bigg[
\alpha +\beta |\Theta|^2+ \gamma \langle \f\rangle^2+
\delta \sum_{ij}O_{ij}^2\bigg] \bigg), 
\end{eqnarray}
%\end{widetext} 
where $\langle \f\rangle=\xi^\dag \f\xi$,   
$O_{ij}=\langle S_i S_j\rangle=\xi^\dag S_i S_j\xi$,  
and
$\Theta=2\xi_3\xi_{-3}-2\xi_2
\xi_{-2}+2\xi_1\xi_{-1}-\xi_0^2$.

We assume that in the ground state $\xi$ is  
position independent. This is equivalent with the single mode approximation, 
where the orbital state of each spin component is taken to be the same. 
This is a realistic assumption if the external potential $U$ is the 
same for all spin components, which is the case in an optical trap. 
The ground-state spinors are then determined by the interaction energy, 
which is proportional to   
\begin{eqnarray}\label{E'}
E'[\psi]&=&\frac{\beta}{|\gamma|} |\Theta|^2+ \frac{\gamma}{|\gamma|}
\langle \mathbf{S}\rangle^2+\frac{\delta}{|\gamma|}O^2\nonumber\\*
&\equiv & b|\Theta|^2+\frac{\gamma}{|\gamma|}\langle\mathbf{S} \rangle^2+dO^2. 
\end{eqnarray}
This is obtained by taking the terms in square brackets in Eq. (\ref{E}),  
dropping the constant $\alpha$ and dividing the resulting equation by $|\gamma|$.  
Following the notation of \cite{Santos05} we have defined $O^2=\sum_{ij}O^2_{ij}$. 
In the energy $0\leq|\Theta|\leq 1$ and  $46\leq O^2\leq 85.5$. The lower 
bound of $O^2$ is achieved for example when $\xi=(0,1,0,0,0,0,0)$ and  
the upper bound when $\xi=(1,0,0,0,0,0,1)/\sqrt{2}$.

Above we have assumed that there is no external magnetic field present. 
If condensate of spin-3 atoms is exposed to a homogeneous magnetic field given by 
$\mathbf{B}=B\mathbf{e}_z$, to first order in the magnetic field the energy changes by  
$E_B[\psi]=-\int d\rr\, p n(\rr) \langle S_z\rangle$. 
Here $p=g\mu_B B$, $\mu_B$ is the Bohr magneton, and $g$ is a constant specific to the condensate atoms.  
We assume that the external 
field is such that it is enough to include this first-order contribution only.       
Since in ${}^{52}$Cr the nuclear spin is zero, for chromium this term 
gives exactly the energy related to the magnetic field. 
In the presence of an external magnetic field the conservation of magnetization 
has to be taken into account. 
The magnetization is defined as 
$\mathbf{M}=\int d\rr\, n(\rr)\langle \mathbf{S}\rangle$, where in the spin operator $\hbar=1$. 
If only the contact interaction is considered, the projection 
of magnetization in the direction 
of the magnetic field is a conserved quantity. 
Thus now the relevant quantity is 
$M_z=\int d\rr\, n(\rr)\langle S_z\rangle$.

We calculate the ground states under the assumptions that the magnetization 
is fixed and the spinor is position independent.   
The $z$ component of magnetization is then determined by $m\equiv\langle S_z\rangle=M_z/N$, where $N$ is the particle number,  
and $E_B[\psi]=-pNm=-pM_z$. 
From this on we call $m$ the magnetization. 
Since $\langle S_z\rangle=m$ is constant, we can drop the term proportional to $\langle S_z\rangle^2$ from the energy, 
which then becomes  
\begin{equation}\label{Em}
E_{m}[\psi]\equiv 
b |\Theta|^2+\frac{\gamma}{|\gamma|}(\langle S_x\rangle^2+\langle S_y\rangle^2)+dO^2. 
\end{equation}

\section{\label{Sec3}GROUND STATES AND PHASE DIAGRAMS}
\subsection{Free magnetization}
We minimize Eq. (\ref{E'}) numerically to 
find out the zero-field ground-state spinors. This minimization has been performed also in Ref. 
\cite{Diener05}, but using an alternative parametrization of the energy. 
Therefore our figures are not directly comparable with those of Ref. \cite{Diener05}. 
We have, however, checked that the zero-field phase diagrams agree exactly  
when the different parametrization is taken into account.   
The paramterization we use makes it easier to see from the phase diagrams 
how the ground states depend on $b$ and $d$. 
The notation of the ground states follows that of Ref. \cite{Diener05}. 

\squeezetable
\begin{table*}[ht]
\caption{\label{Tab2}Here are the ground states. 
Here $m=\langle S_z\rangle$ and $c$ $(r)$ is a complex (real) number. $R$ is a spin rotation given by 
$R(\epsilon_1,\epsilon_2,\epsilon_3)= e^{-i\epsilon_1 S_z}
e^{-i\epsilon_2 S_y}e^{-i\epsilon_3 S_z}$. The value of $\epsilon_2$ is determined by $m$.  
The energy does not change if
 $m_S\rightarrow -m_S$, $S=-3,-2,\ldots,3$, but the sign of magnetization 
changes. In the second and third column we give the notation used in Refs. \cite{Diener05} and \cite{Santos05}, 
respectively. In spinors $a,m$ are such that the terms inside 
square roots are non-negative and $a$ is a function of $b,d$, and $m$. 
}
\begin{ruledtabular}
\begin{tabular}{|c|c|c|c|c|c|c|}
Phase &In \cite{Diener05} & In \cite{Santos05} & $\xi$ & $|\Theta|^2$ & $\langle \mathbf{S}\rangle^2$ & $O^2$ \\
\hline
$A$ &$A$& $P$ &
$(1,0,0,0,0,0,1)$ &  
$1$ & $0$ & $85.5$   \\ 
\hline
$B$ &$B$ & $S_{-3,-1,1,3}$ & $(r,0,c,0,c,0,r)$ &$0$-$1$&$0$&$56$-$85.5$\\
\hline
$C$ &$C$ &$CY_{-3,-1,1,3}$ & $(r_3,0,r_1,0,r_{-1},0,r_{-3})$& $<0.1 $& $< 0.1$ & $> 55$\\
\hline
$D$ &$D$&& 
$(0,1,0,0,0,1,0)/\sqrt{2}$& $1$ & $0$ & $48$\\
\hline
$E$ &$E$ & & $(a,0,0,\sqrt{1-2a^2},0,0,a)$& $\leq 1/81$ & $0$ & $48$-$48.375$\\
\hline 
$F$ & $F$& & $(0,1,0,0,0,0,0)$& $0$& $4$ & $46$\\
\hline
$FF$ & $FF$ &$F$ & $(1,0,0,0,0,0,0)$ & $0$&$9$& $81$\\
\hline
$G$ & $G$ & $S_{-2,0,2}$ & $(0,r_2,0,r_0,0,r_{-2},0)$ &$0$-$1$ &$0$-$4$&$66.5$-$85.5$\\
\hline
$A_m$ &$A_1$& $P$ &
$(\sqrt{\frac{m+3}{6}},0,0,0,0,0,\sqrt{\frac{3-m}{6}})$ &  
$1-\frac{1}{9} m^2$ & $m^2$ & $85.5-\frac{1}{2}m^2$   \\ 
\hline
$B_m$ & $B_1$ &&  $(r_3,0,c_1,0,c_{-1},0,r_{-3})$& $0$-$1 $& $m^2$ & $46.9068$-$85.5$\\
\hline
$C_m$ &$C_1$&& $(r_3,0,r_1,0,r_{-1},0,r_{-3})$& $0$-$1$& $m^2$ & $46.9068$-$85.5$\\
\hline
$D_m$ &&& 
$(0,\sqrt{2+m},0,0,0,\sqrt{2-m},0)/2$& $1-\frac{1}{4}m^2$ & $m^2$ & $48-\frac{1}{2}m^2$\\
\hline
$E_m$&&& $(r_3,r_2,r_1,r_0,r_{-1},r_{-2},r_{-3})$ &  $0$ &$m^2$ & $48-\frac{1}{2}m^2$\\
$E_m^2$&&& $(r_3,r_2,r_1,r_0,r_{-1},r_{-2},r_{-3})$ &$0$ & $m^2$& $18+7m^2$\\
\hline
$F_m^R$ &&& $R\cdot F$& $0$& $4$ & $46$\\
\hline
$FF_m^R$ &&&$R\cdot FF$ & $0$&$9$& $81$\\
\hline
$G_m$ &$G_1$&& $(0,\sqrt{a^2+\frac{1}{4}m},0,\sqrt{1-2 a^2},0,\sqrt{a^2-\frac{1}{4}m},0)$ &$0$-$1$ &$m^2$&$46$-$85.5$\\
\hline
$H_m$ &$H_1$&$CY_{-3,2},CY_{-1,2}$ & $(0,\sqrt{\frac{3+m}{5}},0,0,0,0,\sqrt{\frac{2-m}{5}}),\,\,
(0,0,\sqrt{\frac{2+m}{3}},0,0,\sqrt{\frac{1-m}{3}},0)$   
& $0$ &$m^2$ & $45+(3+|m|)^2$ \\
\hline
$H_m^R$ &&& $R\cdot(a,0,0,0,0,\sqrt{1-a^2},0),\,\, R\cdot (0,a,0,0,\sqrt{1-a^2},0,0)$ & $0$&  
$\left(\frac{3d}{d+1}\right)^2 >m^2$  & $45+\left(\frac{3}{d+1}\right)^2$ \\
\hline
$Z_m$ &$Z_1,Z_2,Z_3$&& $(c_3,c_2,c_1,r_0,c_{-1},c_{-2},c_{-3})$ & $0$-$1$ & $0-9$ & $%54 
46$-$85.5$\\
\hline
$Z_m^R$ &&& $(c_3,c_2,c_1,r_0,c_{-1},c_{-2},c_{-3})$ & $0$-$1$ & $>m^2$ & $46$-$85.5$\\
\hline
$J_m$ &&& 
$(\sqrt{a^2+\frac{m}{6}},0,0,\sqrt{1-2a^2},0,0,
-\sqrt{a^2-\frac{m}{6}})$ & $0$-$1$ & $m^2$& $46$-$85.5$ \\
\end{tabular}
\end{ruledtabular}
\end{table*}
\begin{figure}[h]
\includegraphics[scale=1]{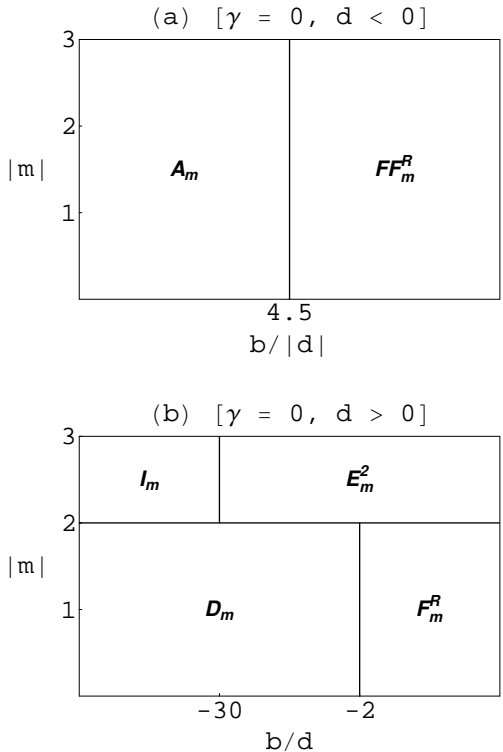}
\caption{\label{Figg0}The ground-state phase diagrams if $\gamma=0$. In 
(a) for $d<0$ and in (b) for $d>0$.    
}  
\end{figure}
\begin{figure*}[h]
\centering 
\includegraphics[scale=1]{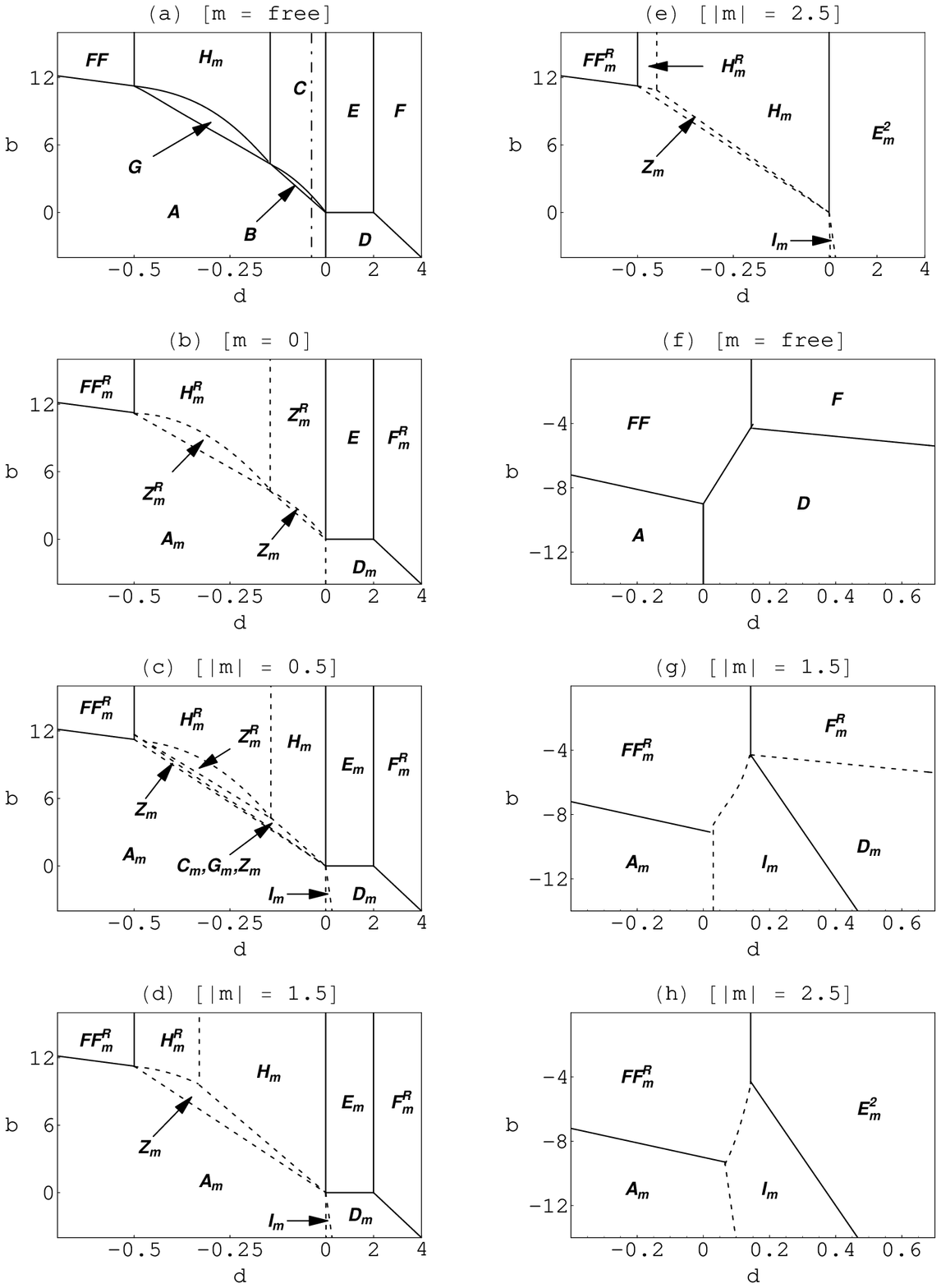}
\caption{ \label{Fig}The ground-state phase diagrams. In (a)--(e) we have $\gamma>0$ and in (f)--(h) $\gamma<0$. 
Here $b=\beta/|\gamma|$ and $d=\delta/|\gamma|$. 
Notice that in (a)--(e) the scale of $d$-coordinate changes at $d=0$. 
In (a) the dot-dashed line is determined by the 
measured values of the scattering lengths of ${}^{52}$Cr and it is located at $d=-0.037$. 
It intersects the $A$-$B$ boundary at $b=1.11$ and 
the $B$-$C$ boundary at $b=1.53$, which correspond to $a_0=9.2a_B$ and $a_0=28.7 a_B$, respectively. 
In all figures except (a) and (f) the dashed line means that 
the phase boundary depends on the value of magnetization, while the  
solid line denotes that the boundary is independent of magnetization.  
} 
\end{figure*}

Next we briefly discuss the ground-state phases.  
If $\gamma>0$ it is in general favorable to have $\langle\mathbf{S}\rangle=0$, whereas if 
$\gamma <0$ it is energetically advantageous to make $|\langle\mathbf{S}\rangle|$ as large as possible. 
This is why the $F$ and $FF$ phases are favored in the figures 
correponding to $\gamma<0$.                     
If $b,d<0$, the energy is minimized when $|\Theta|$ and $O^2$ have their maximum values,   
 i.e., $|\Theta|=1$, $O^2=85.5$ and the system is in $A$-phase. 
If $d<0$ and $b$ is positive it is favorable to have $O^2=85.5$ and  $|\Theta|=0$. 
However, these can not be achieved simultaneously. If $\gamma>0$, this leads to competition 
between several different phases, whereas if $\gamma<0$, $FF$ is the ground state.   
If $b,d>0$ it is favorable to minimize $O^2$ and $|\Theta|$. This achieved if the 
ground state is $F$, for which $O^2=46$ and $\Theta=0$. If $\gamma>0$ $F$ is replaced 
by $E$ if $d$ is small, since the latter has $\langle \mathbf{S}\rangle=\mathbf{0}$. 
Finally, if $b<0$, $d>0$, it is favorable to set $|\Theta|=1$ and $O^2=46$. 
These cannot be achieved simultaneously, and the ground state is $D$, for which 
$|\Theta|=1, O^2=48$ and $\langle\mathbf{S}\rangle=\mathbf{0}$.

\subsection{Fixed magnetization}
  
We have minimized the energy shown in Eq. (\ref{Em}) using $m=0,0.5,1.5$ and $2.5$ as the values of magnetization. 
The ground states are shown in Table \ref{Tab2}. Because many of the ground-state spinors  
 resemble the zero-field ground states, we have used the same letters to label them, 
but have added the subscript $m$ which refers to a fixed magnetization. 
The ground states depend on the signs of $b,d$ in a similar fashion than in the  case of free magnetization.
In most phases $\langle\mathbf{S}\rangle$ is parallel to the magnetic field, i.e.,  $\langle\mathbf{S}\rangle
=\langle S_z\rangle\mathbf{e}_z$. The exceptions are $F_m^R,FF_m^R,H_m^R$, and $Z_m^R$ phases, 
for which $\langle S_x\rangle^2+\langle S_y\rangle^2> 0$. These states cannot be ground states  
if the magnetization is free, since then it is favorable to rotate the spin so that in the ground state  
$\langle\mathbf{S}\rangle$ is parallel to the magnetic field. 
An explicit form for $F_m^R,FF_m^R$ and $H_m^R$ 
can be obtained with the help of the spin rotation matrix $R(\epsilon_1,\epsilon_2,\epsilon_3)=e^{-i\epsilon_1 S_z}
e^{-i\epsilon_2 S_y}e^{-i\epsilon_3 S_z}$. The 
value of $\epsilon_2$ is fixed by the values of $d$ and $m$, whereas $\epsilon_1$ and 
$\epsilon_3$ are free. For example for the first $H_m^R$  
spinor in Table \ref{Tab2} $\epsilon_2=\text{arccos}[-(d+1)m/3d]$ or  $\epsilon_2=\text{arccos}[(d+1)m/3d]+\pi$.  
These result in different spinors. For $FF_m^R$ spinor $\epsilon_2=\text{arccos}(m/3)$ and for 
$F_m^R$  $\epsilon_2=\text{arccos}(m/2)$. 
In $Z_m$ and $Z_m^R$ spinors all spin components are populated and the exact form 
of the spinor depends on $b$ and $d$.  

The energy of a condensate does not depend on the sign of $m$, only on the magnitude. 
Hence phase diagrams with $m=-0.5,-1.5,-2.5$ are similar to those with $m=0.5,1.5,2.5$, respectively. 
The phase diagrams for the special case $\gamma=0$ are shown in Fig.~\ref{Figg0}, while those  
 for positive and negative $\gamma$ are shown in Fig.~\ref{Fig}. The (a)--(e) diagrams in Fig.~\ref{Fig} correspond to 
$\gamma >0$, whereas those for $\gamma<0$ are  shown in (f)--(h).   
If $\gamma<0$ the diagram with free magnetization and that with $m=0$ are similar 
and  the differences between $|m|=0.5$ and $|m|=1.5$ diagrams are very small, 
so we have omitted the $|m|=0$ and $|m|=0.5$ diagrams. 
The phase diagrams change at $m=2$ because $D_m,E_m$, and $F_m^R$ are not possible ground states if $m>2$, 
but are  replaced by $E_m^2$.  

In most ground states 
$\langle S_x\rangle=\langle S_y\rangle=0$, so the energy becomes    
\begin{equation}\label{Emp}
E_{m}[\psi]= b|\Theta|^2+dO^2.   
\end{equation}
The ground state of this equation is determined by the ratio $b/d$ and the sign of 
$d$. If we multiply $b$ and  $d$ by a positive constant 
the ground state is unchanged.  
Equation (\ref{Emp}) shows that 
if $(d,b)$ is on the boundary of two phases which both have $\langle S_x\rangle=\langle S_y\rangle=0$ 
and $x$ is a positive number, then 
also $(xd,xb)$ is on the same boundary.  
Therefore the phase boundaries between phases with  $\langle S_x\rangle=\langle S_y\rangle=0$, 
$\langle S_z\rangle=m$ are straight lines with one end at the origin.  
This reasoning does not work if $\langle S_z\rangle$ depends on $b$ or $d$, like in $B$ and $G$ phases.

On a phase boundary the energies of two phases, labeled now by $1$ and $2$, are equal.  
Then Eq. (\ref{Emp}) gives  $b=-d(O_1^2-O_2^2)/(|\Theta_1|^2-|\Theta_2|^2)$ and by using the values 
for $O^2$ and $|\Theta|^2$ listed in Table \ref{Tab2}
equations for some of the phase boundaries can be obtained. 
Next we discuss the phase diagrams in more detail.

\subsubsection{$\gamma>0$}
In Fig.~\ref{Fig}(a) the boundary between $C$ and $H_m$ is at $d=-0.15$, which corresponds to   
$\langle S_z\rangle^2=0.1$. In the $H_m$ spinor $m$ depends on $d$ according to $m=-3d/(1+d)$. 
 $A$ and $B$ are separated by $b=-30d$, while the 
$B$-$C$ boundary is $b=-43.83 d-98.78 d^2$. 
$A$-$G$ and $H_m$-$G$ boundaries are given by $b=1.5-19.5 d$ and 
$b=-3.20-63.01 d-76.97 d^2+34.82 d^4$, respectively.  
$FF$ and $A$ are separated by $b=9-4.5d$ and 
$F$ and $D$ by $b=4-2d$.  

In Fig.~\ref{Fig}(b) the boundary 
between $H_m^R$ and $Z_m^R$ is located at $d\approx -0.14$. 

In Figs.~\ref{Fig}(c)--\ref{Fig}(e)  
$H_m^R$ and $H_m$ are divided by a vertical line located at 
$d=-|m|/(3+|m|)$. This equation is valid if $|m|\geq 0.5$, 
for smaller $|m|$ $H_m$ is replaced by $Z_m^R$.  
The value of $a$ in $H_m^R$ is $a=\sqrt{(2-d)/5(d+1)}$ in the first spinor in Table \ref{Tab2} and 
$a=\sqrt{(1+4d)/3(1+d)}$ in the second one. These are defined only for those values of $d$ for which 
the term inside the square root is non-negative.
The boundary between $A_m$ and $Z_m$ is 
given by $b=-22.5d$. The  $Z_m$-$H_m$ boundary depends on $m$ and is 
given approximately by $b=(-36.25+4.71|m|)d$ 
for $|m|\in [1,3]$. $I_m$ and $D_m$ are separated by $b=-30 d$. 

\subsubsection{$\gamma<0$}
It can be seen that if $\gamma<0$ the phase diagrams depend only weakly 
on the value of $|m|$. The ferromagnetic states $F_m^R$ and $FF_m^R$ are favored 
since they allow $\langle\mathbf{S}\rangle^2$ to have large values. 
In Fig.~\ref{Fig}(f) $F$ and $D$ are separated by $b=-2-4 d$. 
$FF$ and $F$ meet at $d=1/7$ and 
$FF$-$A$ boundary is given by $b=-9-4.5d$. In Fig.~\ref{Fig}(g) $D_m$ and $I_m$ are separated by $b=-30 d$. 

\subsubsection{$\gamma=0$}
The phase diagrams corresponding to $\gamma=0$ are shown in Fig.~\ref{Figg0}. 
There are two phase diagrams, in (a) for $d<0$ and in (b) for $d>0$. Since $\gamma=0$, 
there are two free parameters, the ratio $b/|d|$ and magnetization $m$. 
The phase diagram in Fig.~\ref{Figg0}(a) can be obtained from those of Fig.~\ref{Fig} 
by taking the limit $d\rightarrow -\infty$ with $b/d$ fixed. In this limit the term $\langle S_x\rangle^2+\langle S_y\rangle^2$ in 
Eq. (\ref{Em}) becomes negligible. If $\gamma\not=0$, the boundary 
between $A_m$ and $FF_m$ is at $b=-4.5d + 9\gamma/|\gamma|$. If $d\rightarrow-\infty$ with $b/d$ fixed, 
this equation gives $b/|d|=4.5$, which is the phase boundary in Fig.~\ref{Figg0}(a). In the same way, by letting 
$d\rightarrow\infty$ with $b/d$ fixed, one obtains the phase diagram of Fig.~\ref{Figg0}(b).

\subsection{Ground states of ${}^{52}$Cr}
\begin{figure}[h]
\includegraphics[scale=1]{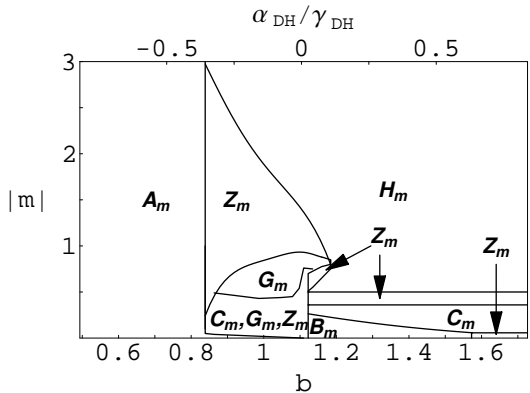}
\caption{ \label{FigCr}The ground states of ${}^{52}$Cr. At the top of the figure
we give the scale used in Fig. 5 of Ref. \cite{Diener05} in order to make a comparison with their phase diagram easier.  
Here $\alpha_{DH}/\gamma_{DH}=1.35\, b-1.48$. 
The relation between $b$ and the unknown scattering length $a_0$ is given by $a_0=-42.15+46.28\, b$, 
where $a_0$ is given in units of Bohr radius.      
} 
\end{figure}
In Fig.~\ref{FigCr} we have plotted the ground states of ${}^{52}$Cr 
as a function of $b$ and $|m|$. In this figure $d=-0.037$, which is the 
value relevant for chromium. As explained above,  
phase boundaries between states with $\langle S_x\rangle=\langle S_y\rangle=0,\langle S_z\rangle=m$ are 
of the form $b=kd$, where $k$ is a constant specific to the boundary. If $|m|,\gamma>0$ and $-0.14<d < 0$ one 
has $\langle S_x\rangle=\langle S_y\rangle=0$. 
Therefore Fig.~\ref{FigCr} holds true for every $d'$ in the interval $(-0.14,0)$, but with 
$b$ replaced by $b'=b\, d'/d_{Cr}$,
where $b$ is the coordinate used in Fig.~\ref{FigCr} and $d_{Cr}=-0.037$. 
%For chromium $d=-0.037$ and $b=0.911+0.022a_0/a_B$.
If we assume that the uncertainty in the values of $a_2,a_4,a_6$ is $\pm 10a_B$,   
then $d\in [-0.08,0]$. Thus Fig.~\ref{FigCr} is valid for ${}^{52}$Cr 
also if a small uncertainty in the values of scattering lengths is allowed. 

It is interesting to note that Fig.~\ref{FigCr} closely resembles the Fig. 5 of Ref. \cite{Diener05}. 
The latter figure gives the ground state of ${}^{52}$Cr as a function of $b$ and external magnetic field,
calculated allowing the magnetization to vary freely. 
The similarity is a consequence of the fact that in the figure of Ref. \cite{Diener05} the magnetization 
of the ground state can approximately be written as $m=cB$, where $B$ is the strength of the 
magnetic field and $c$ is a constant. 
In general, however, the relation between the strength of an external magnetic field and the magnetization 
of the corresponding ground state is not of this form.    
Therefore the phase diagrams in Fig.~\ref{Fig} are not identical with phase diagrams  
showing the ground state in some given magnetic field as a function of $b$ and $d$.

\section{\label{Sec4}DIPOLE-DIPOLE INTERACTION AND THE GROUND STATES OF ${}^{52}$Cr}
\subsection{Dipole-dipole interaction}
Above we have neglected the magnetic dipole-dipole interaction.  
If $\xi$ is position independent, 
the dipole-dipole interaction energy is 
\begin{equation}\label{Edd}
E_{dd}[\psi]=\frac{c_{dd}\langle \mathbf{S}\rangle^2}{2}\int d\rr\, n(\rr)\int d\rr'\frac{n(\rr')[1
 -3(\mathbf{\hat{x}}\cdot\langle\mathbf{\hat{S}}\rangle)^2]}{|\rr-\rr'|^3}, 
\end{equation}
where $\langle\mathbf{\hat{S}}\rangle=\langle\mathbf{S}\rangle/|\langle\mathbf{S}\rangle|$, 
$\mathbf{\hat{x}}=(\rr-\rr')/|\rr-\rr'|$. For ${}^{52}$Cr $c_{dd}=\mu_0 (g_e \mu_B)^2/4\pi$, where  
$\mu_0$ is the vacuum permeability, $\mu_B$ is the Bohr magneton and $g_e\approx 2$.

The effect of the dipole-dipole interaction is to change 
the value of $\gamma$ and possibly the direction of $\langle\mathbf{S}\rangle$. 
In most experiments the trap is weak in one direction and strong in 
orthogonal directions, which produces a cigar-shaped condensate.  
If $\langle\mathbf{S}\rangle$ is parallel to the long axis of the trap, 
the cloud remains cigar shaped also after the 
 introduction of the magnetic dipole-dipole interaction \cite{Odell04,Eberlein05}. 
Based on Eq. (\ref{Edd}) one can argue that in a cigar-shaped condensate the minimum of $E_{dd}$  
is obtained when the spin is everywhere parallel to the long axis of the condensate. 
Thus in the absence of an external magnetic field the dipole-dipole interaction aligns the 
spin with the long axis of the condensate \cite{Diener05}.

The assumption of a position independent spinor as the 
ground state is not always realistic. In a ${}^{87}$Rb condensate of spin-2 atoms 
the quadratic Zeeman term may favor ground states consisting of regions in different spin states. 
The ground state with $M_z=0$ may, for example, consist of regions where the projection of spin is  either 
$m_S=2$ or $m_S=-2$, and the total magnetization adds up to $0$ \cite{Saito05}. 
This kind of effect can also be caused by the magnetic dipole-dipole 
interaction. Let us assume that the long axis of a cigar-shaped condensate and magnetic field  are parallel.    
In most ground states $\langle\mathbf{S}\rangle$ is parallel to the magnetic field, which is the direction that  
minimizes the dipole-dipole interaction energy.  
However, for $F_m^R,FF_m^R,H_m^R$, and $Z_m^R$ states this is not true, 
and therefore the dipole-dipole interaction energy is not minimized. 
The minimum of energy is obtained by forming regions where the 
spin is either along the positive or negative $z$ axis.  The magnetization of these regions has to 
add up to $M_z$. For example for $FF_m^R$ state the regions would be either in $m_S=3$ or $m_S=-3$ state.    
Forming these regions costs some kinetic energy, which has to be overwhelmed by 
the energy released from the dipole-dipole interaction.

\subsection{Ground states of ${}^{52}$Cr}
The scattering lengths of ${}^{52}$Cr indicate that in the absence of 
dipole-dipole interaction and for free magnetization the ground state is 
$A$, $B$, or $C$. If the magnetization is fixed and $|m|$ is larger than $0.5$, 
 the ground state is either $A_m$, $G_m$, $H_m$, or $Z_m$. 
The difference between these states is easy to see experimentally, 
since they have different spin components populated.

The introduction of the dipole-dipole interaction may change the ground states. 
For chromium the value of $\gamma$ is positive and $\delta$ is negative. 
We assume that the change in $\gamma$ following from the  
dipole-dipole interaction is such that $\gamma$ remains positive.  
A change in $\gamma$ shifts 
the point $(\frac{\delta}{|\gamma|},\frac{\beta}{|\gamma|})$ characterizing chromium 
radially toward or away from the origin. It, however, cannot cross the origin.

If magnetization is free, one can see from Fig.~\ref{Fig} that the $H_m$ phase is reached if $\delta/\gamma<-0.15$   
and $b$ is large enough. 
For ${}^{52}$Cr these conditions are fulfilled if the change $\Delta\gamma=-|\Delta\gamma|$ induced by the dipole-dipole 
interaction is such that $|\Delta\gamma|/\gamma\geq 0.76$ and $b>1.18$. The latter is equivalent with $a_0>12.6a_B$.

If magnetization is fixed,     
the ground state is either unchanged or becomes $H_m^R,FF_m^R$, or $Z_m^R$ as the value of $\gamma$ changes.  
 The amount of change in $\gamma$ required to reach $H_m$ depends on the 
value of magnetization. If $|m|\leq 0.5$, the boundary between $Z^R_m$ and $H_m^R$ is at 
$d\approx -0.14$, which requires $|\Delta\gamma|/\gamma \geq 0.74$.  
For $|m|>0.5$ the $H_m$-$H_m^R$ boundary is at $d=-|m|/(3+|m|)$, which shows  that 
 $|\Delta\gamma|/\gamma \geq 0.96-0.11/|m|\geq 0.74$ is required in order to reach the  
$H_m^R$ state. Also the minimum value of $b$ for this to be possible is a function of $|m|$, but 
if $b>1.18$, or $a_0>12.6a_B$, $H_m$ can be reached, regardless of the value of $m$.

\begin{figure}[h]
\includegraphics[scale=1]{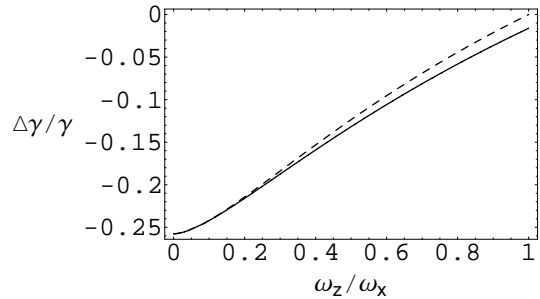}
\caption{ \label{DG} The change in $\gamma$ induced by the dipole-dipole interaction. 
Solid line is for $|m|=3$ and dashed line for $m=0$. These give the maximum and minimum for 
$|\Delta\gamma|$, respectively. If $m=0$, the ground state is independent of $\Delta\gamma$.       
} 
\end{figure}

The change in $\gamma$ caused by the magnetic dipole-dipole interaction 
can be estimated by using the analytical results for the dipole-dipole and contact 
interaction energies obtained in \cite{Odell04,Eberlein05}. These results hold in the Thomas-Fermi limit, 
where the kinetic energy is negligibly small compared with the interaction and trapping energies.  
We assume that $\langle\mathbf{S}\rangle=m\mathbf{e}_z$, 
the trapping potential is cylindrically symmetric, $U(\mathbf{r})=M[\omega_x^2 (x^2+y^2)+\omega_z^2 z^2]/2$, 
and that $\omega_z/\omega_x<1$, so that the trap is cigarlike. In Fig. \ref{DG} we plot the ratio 
$\Delta\gamma/\gamma$ as a function of the trap aspect ratio  $\omega_z/\omega_x$.  
The minimum value is $\Delta\gamma/\gamma\approx -0.26$. A similar result was found in 
\cite{Diener05} using a Gaussian ansatz for the particle density. 
In addition to the dipole-dipole interaction, an uncertainty in the values of the scattering lengths may produce a  
negative $\Delta\gamma$. If  the values 
of $a_2,a_4$, and $a_6$ are known within   
$\pm 10a_B$, the maximum change is $|\Delta\gamma|/\gamma=0.17$.  
Thus the change in $\gamma$ resulting from the dipole-dipole interaction and a possible uncertainty in the scattering lengths 
does not appear to be big enough to change the ground state of a chromium condensate. 

\subsection{Experiments with ${}^{52}$Cr}

In a high magnetic field the coherent spin relaxation via magnetic dipole-dipole interaction 
is suppressed. This occurs when the energy spacing of the neighboring Zeeman levels, $g_e\mu_B B$,  
is much larger than the chemical potential $\mu$. For a chromium condensate of $10^5$ particles 
these energies are equal when $B\sim 2$ mG. 
Although coherent spin relaxation is prevented, spin relaxation can still occur via incoherent processes. In these 
the energy released in spin relaxation is so large that the colliding 
particles escape from the trap. 
For ${}^{52}$Cr this process has been studied both theoretically and experimentally in Ref. \cite{Hensler03}. 
For example for $m_S=3$ state at a magnetic field of 
$1$ G the rate constant $\beta_{dr}$ characterizing inelastic  dipolar relaxation was found to be 
$\beta_{dr}=4\times 10^{-12} $ cm${}^3$ s${}^{-1}$ \cite{Hensler03}. Thus 
for a typical central density $n=10^{14} $ cm${}^{-3}$ the $m_S=3$ state decays  
in a few ms. This is a short time compared with the typical timescales 
of ${}^{52}$Cr experiments, which are on the order of tens of ms \cite{Stuhler05,Giovanazzi06c}. 
One way to 
suppress the incoherent dipolar relaxation is to reduce the condensate density. This can be done 
by loading the condensate to an optical lattice of shallow traps \cite{Takamoto05}. If one  
reduces the density, e.g., by two orders of magnitude, the new decay time is on the order of   
hundreds of milliseconds, which should be long enough for experiments. 

From experimental point of view it is important to be able to produce 
different spinors as initial states. This can be done using rapid adiabatic passage and controlled Landau-Zener 
crossing techniques, as described in Ref. \cite{Mewes97}. For spin-1 and spin-2 ${}^{87}$Rb condensates this 
has been done in Ref. \cite{Schmaljohann04}.
For $|m|>0.5$, the ground state of  ${}^{52}$Cr is 
very likely of the form $(0,\sqrt{(3+m)/5},0,0,0,0,\sqrt{(2-m)/5})$. 
If a condensate is initially prepared in this state, it should not show any coherent spin dynamics. 
Because of incoherent processes the populaton of the $m_S=2$ component decreases, 
while that of $m_S=-3$ component is almost unchanged. This leads to a decrease of magnetization. 
If initially $m<-0.5$, at every later moment the condensate is in the instantaneous ground state 
of spin with magnetization $m<-0.5$.

\section{\label{Sec5}CONCLUSIONS}
We have calculated the ground states and the ground-state phase diagrams of spin-3 BECs 
using mean-field theory and single-mode approximation. 
We have assumed that the projection of the magnetization in the direction of the external  
magnetic field is a conserved quantity. 
We have presented the ground-state phase diagrams pertaining to several values of magnetization.  
The phase diagrams are classified by the value of $\gamma$, 
which is determined by the scattering lengths of different scattering channels. 
We found out that the phase diagrams with $\gamma<0$ depend only weakly on the 
value of magnetization, whereas if $\gamma>0$ the dependence is stronger.  
The phase diagrams for a $\gamma=0$ condensate are obtained in a certain limit 
from the $\gamma>0$ or the $\gamma<0$ diagrams. 

We have plotted the ground states of ${}^{52}$Cr as a function of magnetization and 
the unknown scattering length $a_0$. We found out that this resembles the phase diagram 
for a chromium condensate in a weak magnetic field, where the magnetization can vary freely.  
This follows from the fact that in the latter the magnetization of the ground state 
depends approximately linearly on the strength of the magnetic field. 

We have also studied how 
the magnetic dipole-dipole interaction affects the ground states. 
In the absence of the dipole-dipole interaction it is possible to have ground states where 
the magnetization is not parallel to the magnetic field.  
We have showed that, instead of these states, the dipole-dipole interaction may favor ground states where the condensate 
 has broken up into regions having the magnetization parallel to the magnetic field. 
We have illustrated that the ground state of ${}^{52}$Cr  
does not appear to depend on whether or not the contribution from the dipole-dipole interaction is included. 
Finally we have discussed the experimental realization of spinor ${}^{52}$Cr condensates, pointing  
out that due to fast incoherent spin relaxation the condensate density should be decreased in order to make 
the lifetime of condensate long enough for experiments.

\begin{acknowledgments}
The authors acknowledge the financial support 
of the Academy of Finland (Grants No. 206108 and No. 115682). H.M. was supported by the 
Finnish Academy of Science and 
Letters, Vilho, Yrj\"o and Kalle V\"ais\"al\"a Foundation.
\end{acknowledgments}

\end{document}